\documentclass{ws-ijwmip}

\usepackage[slantedGreek]{mathptmx}

\newcommand\bw{}  

\newcommand{\beq}{\begin{equation}}
\newcommand{\eeq}{\end{equation}}
\newcommand{\bea}{\begin{eqnarray}}
\newcommand{\eea}{\end{eqnarray}}

\newcommand{\mrm}[1]{\mathrm{#1}}
\newcommand{\wt}[1]{\widetilde{#1}}
\newcommand{\real}{\mathrm{Re}\,}

\begin{document}

\markboth{ROBERT W. JOHNSON}{SYMMETRIZATION AND ENHANCEMENT OF THE MORLET TRANSFORM}

\catchline{}{}{}{}{}

\title{SYMMETRIZATION AND ENHANCEMENT OF THE CONTINUOUS MORLET TRANSFORM FOR SPECTRAL DENSITY ESTIMATION}

\author{ROBERT W. JOHNSON}
\address{Alphawave Research, 29 Stanebrook Court\\
Jonesboro, GA 30238, USA\\
robjohnson@alphawaveresearch.com}


\maketitle

\begin{history}
\received{19 August 2010}
\revised{14 January 2011}
\end{history}

\begin{abstract}
The forward and inverse wavelet transform using the continuous Morlet basis may be symmetrized by using an appropriate normalization factor.  The loss of response due to wavelet truncation is addressed through a renormalization of the wavelet based on power.  The spectral density has physical units which may be related to the squared amplitude of the signal, as do its margins the mean wavelet power and the integrated instant power, giving a quantitative estimate of the power density with temporal resolution.  Deconvolution with the wavelet response matrix reduces the spectral leakage and produces an enhanced wavelet spectrum providing maximum resolution of the harmonic content of a signal.  Applications to data analysis are discussed.
\end{abstract}

\keywords{ {Continuous wavelet transform}; {Wavelet enhancement}; {Power spectral density}; {Data analysis}. }

\ccode{AMS Subject Classification: 42C40, 65T60}


\section{Introduction}
\label{intro}
The continuous wavelet transform using the Morlet basis\cite{Morlet:1984} has become quite popular both for theoretical analysis\cite{kaiser:1990,kaiser:1994,perel-3441P,perel-5211P,lewalle-131,karak-313} and for data analysis.\cite{echera-41,fligge-313,Foster:1996,Frick:1997426,Frick:1997670F,ligao-181,piscaron-1661,polygian-725,sweldens98lifting,torrence:98,weber-917}  There is some variety in the literature as to the assignment of the normalization factors, and we propose a rearrangement so as to produce a symmetric forward and inverse transform pair.  The loss of response due to wavelet truncation known as the cone-of-influence is addressed through a renormalization of the wavelet amplitude which keeps its power constant for a given scale.  The renormalized power spectral density may then be enhanced by deconvolution with the wavelet response matrix, yielding the maximum spectral resolution of the harmonic content.  We conclude by discussing the utility of these algorithms and point out a recent application.

The lack of a quantitative power spectral density has long hampered wider adoption of the continuous wavelet transform for data analysis.  A mathematical engineer wants more than just a pretty picture---being able to give a numerical estimate to the power carried within a particular frequency band is of practical importance, and the use of a wavelet rather than Fourier transform allows that estimate to be time dependent.  When the data is limited in duration, interesting features may be located in that region where wavelet truncation has become a significant effect.  By renormalizing the wavelets for constant power, the useful range may be extended beyond the cone-of-influence, and nearly perfect reconstruction holds for all but the edge-most sample locations.  Furthermore, the power spectral density inherits the units of the signal such that its margins yield physical estimates for the instant power and mean spectral density.

The breadth in scale of the wavelet response to a pure signal tone is a consequence of its localization in phase space, as the spectral and temporal resolutions are inversely related.  Again, what a mathematical engineer wants is a precise identification of the frequency spectrum, in which features are maximally resolved with minimal spectral leakage.  By treating the continuous wavelet transform as a theoretical apparatus acting upon a signal, one may in essence calibrate the device given some basic assumptions on the form of the signal components.  Minimizing the discrepancy between the continuous instant wavelet power and the convolution of the calibration matrix with the enhanced spectral estimate then yields the sharpest resolution of the time-varying harmonic content of a signal.

The continuous wavelet transform differs from the discrete transform in some important ways.  Most notable is the highly redundant nature of the analyzing functions, which do not form an orthonormal basis.  Nonetheless, Plancherel's theorem for energy conservation holds, indicating the continuous transform may be used for quantitative power spectral density estimation.  While the continuous transform must be discretized for numerical evaluation, its resolution in scale is arbitrary, leading to the possibility of spectral enhancement.  In contrast, the discrete wavelet transform selects only those scales which do provide an orthonormal basis and may not be enhanced by the method presented here.

\section{Normalization and Central Frequency}
\label{normandfreq}
We first discuss the symmetrization of the forward and inverse transform and its effect on the central frequency employed.  The time unit throughout this investigation is set by the sample rate $1/\Delta_t \equiv 1$.  One may write the usual Morlet wavelet\cite{Frick:1997426,torrence:98} at scale $s = 1 / f_s = 2 \pi / \omega_s$ and offset $t$ using the parameter $\eta \equiv (t' - t) / s$ as the product of a scale dependent normalizing constant $C$, a unit magnitude Gaussian window $\Phi$, and a unit magnitude Fourier wave $\Theta$, \beq
\psi^0_{s,t}(t') \equiv C^0_s \Phi^0_{s,t}(t') \Theta^0_{s,t}(t') = \pi^{-1/4} s^{-1/2} e^{- \eta^2/2} e^{i \omega_1 \eta} \;,
\eeq where $\omega_1 \approx 2 \pi$ is the central frequency of the mother wavelet at unity scale and zero offset, $\psi^0_{1,0}(t') = \pi^{-1/4} e^{- {t'}^2/2} e^{i \omega_1 t'}$.  The window $\Phi^0_{s,t}(t')$ has a discrete extent of $- \lfloor s \gamma$ to $\lfloor s \gamma$ defined by the parameter $\gamma$, and thus the wavelet $\psi^0_{s,t}(t')$ has a length of $ N_{t'} = 2 \lfloor s \gamma + 1$, where $\gamma = 6$ is used herein; together, $\gamma$ and $\omega_1$ determine the time-scale resolution of the transform.  This mother wavelet is normalized to unit energy so that its Fourier transform $\wt{\psi}^0_{1,0}(\omega') = \int_{-\infty}^\infty \psi^0_{1,0}(t') e^{i \omega' t'} dt' = (2 \pi)^{1/2} e^{-(\omega' - \omega_1)^2/2}$ for $\omega' > 0$ has the integrals $\int_0^\infty \vert \wt{\psi}^0_{1,0}(\omega) \vert^2 d\omega / 2 \pi = [1+\mrm{erf}(\omega_1)]/2 \approx 1$ and $\int_0^\infty \vert \wt{\psi}^0_{1,0}(\omega) \vert^2 d\omega / \omega \approx 1 + \mathcal{O}(10^{-4})$.  The complex wavelet has only positive frequencies in its Fourier spectrum, whereas the spectrum of a real wavelet is reflection symmetric.\cite{kaiser:1994}   The conventional transform pair ({\it cf.} Eqs. (6) and (9) of Ref.~\refcite{Frick:1997426}) of a mean-subtracted signal $y(t) = \sum_k Re (A_k e^{i \omega_k t})$ with duration $N_t$ is written for $s \geq 2$ as \bea
&{\rm CWT}^0(s,t) =& \sum_{t'} \bar{\psi}_{s,t}^0(t') y(t') \;,\\
&{\rm ICWT}^0(t) =& \real \biggl [\sum_s \sum_{t'} \bar{\psi}_{s,t}^0(t')\, {\rm CWT}^0(s,t') \Delta_s / s^2 \biggr ] \;,
\eea which does not exhibit an explicit symmetry in functional form, denoting the complex conjugate as $\bar{\psi}(\eta) = \psi(-\eta)$ for positive scales.

The purpose of the normalizing constant is to equate the wavelet response across scales, and we feel that its form should be the same for the forward and inverse transforms.  Pulling over from the denominator of the inverse transform a factor of the scale $s$ and including a factor $\sqrt{2}$ representing the response at negative scales gives a normalization $C_s = \sqrt{2} \, C^0_s / s$ which produces a transform with some very desirable properties, \beq
\psi_{s,t}(t') \equiv C_s \Phi^0_{s,t}(t') \Theta^0_{s,t}(t') = \sqrt{2} \, \pi^{-1/4} s^{-3/2} e^{- \eta^2/2} e^{i \omega_1 \eta} \;.
\eeq  The mother wavelet now has a squared norm of 2, which we interpret as including the response at negative scales to negative frequencies, and that of a scaled wavelet is now $2/s^2$, noting the factor $\sqrt{2}$ should not be applied when considering the positive and negative scales separately.  (It is the reflection-symmetric form of the Morlet wavelet on the time axis which lets one represent the negative scale response as a constant factor, as an asymmetric wavelet requires separate attention to the positive and negative regions of the scale axis.)  By analogy with the photon, the energy of a localized wave is proportional to its frequency $E_\nu \propto \nu = s_\nu^{-1}$, thus its power (energy per time) should be proportional to its energy over its period, $P_\nu \propto s_\nu^{-2}$.  The forward and inverse transform pair are now formally symmetric, \bea
&{\rm CWT}(s,t) =& \sum_{t'} \bar{\psi}_{s,t}(t') y(t') \;,\\
&{\rm ICWT}(t) =& \real \biggl [\sum_s \sum_{t'} \bar{\psi}_{s,t}(t')\, {\rm CWT}(s,t') \Delta_s \biggr ] \;,
\eea with nearly perfect reconstruction within the cone-of-influence and quantitative agreement between the estimated power and the sum of the squared amplitudes of the signal components.  Use of logarithmic scale spacing requires retention of the factor $\Delta_s$.  The root-mean-square power spectral density ${\rm PSD}(s,t) \equiv \vert {\rm CWT} \vert^2$ is normalized such that the integrated area of an isolated peak in the instant wavelet power ${\rm IWP}_t(s) \equiv {\rm PSD}(s,t)$ returns half the square of the amplitude $A_k$ of the signal component, whose sum gives the signal power $P_\mrm{rms} = \sum_k A_k^2 / 2$.  The margins of the PSD give the mean wavelet power ${\rm MWP}(s) = N_t^{-1} \sum_t {\rm PSD}(s,t)$ and the integrated instant power ${\rm IIP}(t) = \sum_s {\rm PSD}(s,t) \Delta_s$ as summations over the time and scale axes, respectively.

\begin{figure}[t]
\centerline{\psfig{file=\bw 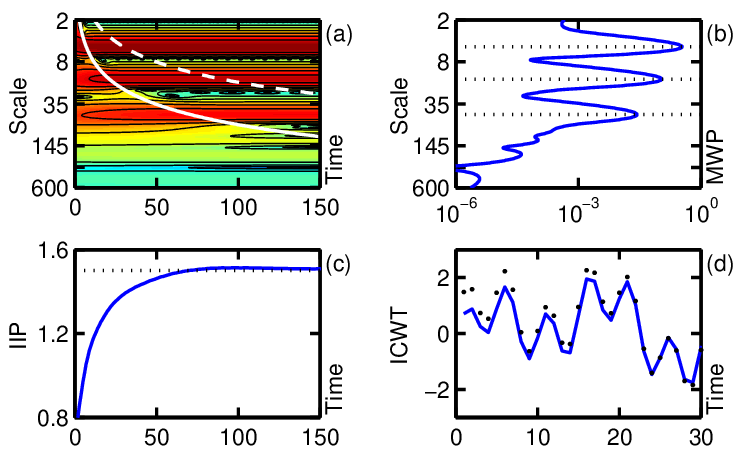}}
\vspace*{8pt}
\caption{\label{fig:A}
CWT power spectral density (a), mean wavelet power (b), integrated instant power (c), and reconstruction (d) for test signal $y_1$ with components of unit amplitude and periods 5, 15, and 50 with rms power of 1.5. An unlabeled tick appears at the scale of the signal duration $N_t=300$. Overlaying the PSD in (a) are the cone-of-influence (solid) and the cone-of-admissibility (dashed).  All abscissas are labeled to their right. }
\end{figure}

The analysis of a test signal $y_1$ of duration $N_t=300$ time units with signal components of unit amplitude and periods of 5, 15, and 50 is shown in Fig.~\ref{fig:A}.  Throughout this paper all abscissas are labeled to the lower right of the plot.  The cone-of-influence defined by the $e$-folding time ($t_e = \sqrt{2} s$ for the Morlet wavelet) is marked with a solid line in (a), and the more restrictive cone-of-admissibility denoting the first wavelet truncation at a given scale is marked with a dashed line.  A trough appears at the scale of the signal duration $N_t$ in (b), beyond which we identify the extremely low frequency (ELF) region where $s > N_t$.  Apparent is the loss of transform response in (c), where the IIP falls below the rms power $P_\mrm{rms} = 1.5$, as is the loss of reconstruction in (d) at the signal edge.

\begin{figure}[t]
\centerline{\psfig{file=\bw 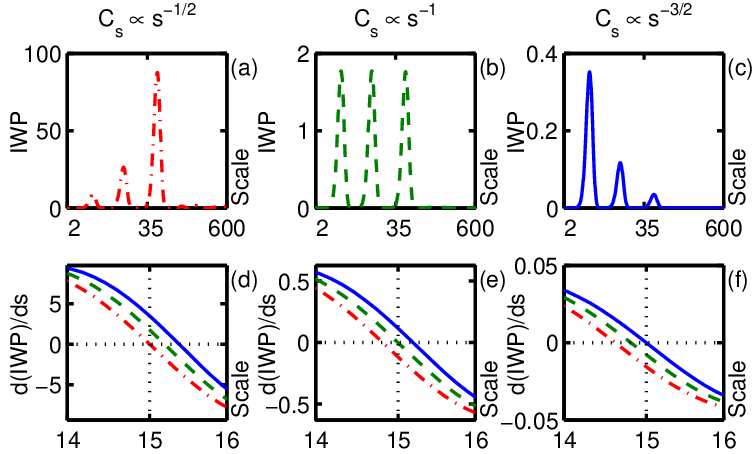}}
\vspace*{8pt}
\caption{\label{fig:B}
Instant wavelet power (top row) and its gradient at the central peak (bottom row) for forward transform normalizations labeled by column and central frequencies of $2 \pi - 1/ 4 \pi$ (dash-dot), $2 \pi$ (dashed), and $2 \pi + 1 \ 4 \pi$ (solid).  Test signal $y_1$ has components of unit amplitude and periods 5, 15, and 50.  Each peak in (c) has an area of 0.5, which is equal to the rms power of the signal component. }
\end{figure}

The central frequency given by Torrence and Compo\cite{torrence:98} to unify the Fourier period and wavelet scale, $\lambda_1 / s_1 = 4 \pi / [\omega_1 + (2 + \omega_1^2)^{1/2}] = 1$ yielding $\omega_1 = 2 \pi - 1 / 4 \pi$, is no longer appropriate for our normalization.  Using the same test signal, we consider transforms with central frequencies $2 \pi - 1 / 4 \pi$, $2 \pi$, and $2 \pi + 1 / 4 \pi$ and forward scalings of $s^{-1/2}$, $s^{-1}$, and $s^{-3/2}$ appearing in the CWT.  The top row in Fig.~\ref{fig:B} displays the instant wavelet power for a single central frequency at the center of the transform $t = N_t/2$, and the bottom row shows its gradient for all three central frequencies in the vicinity of the central signal peak; similar graphs obtain for the other peaks, noting that the forward scaling of $s^{-1}$ in (b) and (e) corresponds to that recently proposed by Liu, {\it et al}.\cite{liu-2093}  Kaiser\cite{kaiser:1994} states that ``the actual value of [the power appearing on the normalization scale] is completely irrelevant to the basic theory'', a position with which we politely disagree.  Only the transform with scaling $s^{-3/2}$ produces peaks with an integrated area equal to half the sum of squared amplitudes, and we note that the locations of its peaks coincide with the signal periods for the central frequency of $\omega_1 = 2 \pi + 1 / 4 \pi$.  The response of the symmetrically normalized CWT is that of a theoretical apparatus whose point spread function preserves the area of a Dirac distribution representing the power carried by a pure signal component of infinite duration with constant amplitude and period.

\section{Renormalization}
\label{renorm}
\begin{figure}[t]
\centerline{\psfig{file=\bw 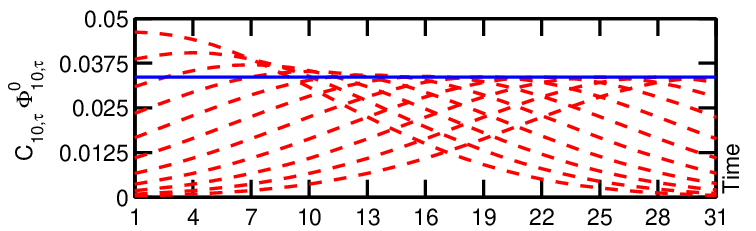}}
\vspace*{8pt}
\caption{\label{fig:H}
Renormalized windows $ C_{s,\tau} \Phi^0_{s,\tau} = \vert \psi_{s,\tau} \vert$ for several offsets at scale $s=10$.  The magnitude within the cone-of-admissibility is indicated by the horizontal line.  The renormalization only becomes significant for $t' \leq \sqrt{2} s$. }
\end{figure}

We next introduce a renormalization which compensates for the reduction in response outside the cone-of-influence.  The cone-of-influence indicates that region beyond which the response of the CWT is significantly affected by the wavelet truncation, which begins at the cone-of-admissibility.  Various algorithms have been proposed for its rectification\cite{Foster:1996,Frick:1997426,sweldens98lifting,liu-2093,rwj:jgr02,rwj:astro01}; however, we have found that algorithms which alter the shape of the analyzing wavelet also affect its frequency response.  Thus, we are led to proposing a simple renormalization such that for transform coefficients outside the cone-of-admissibility the wavelet is given a norm of $(2/s^2)^{1/2}$.  For wavelets truncated by either edge of the signal, the window $\Phi^0_{s,\tau}$ is shifted by an offset $\tau$ relative to an unshifted window $\Phi^0_{s,0}$ defining the time span $t'$.  The length of a truncated wavelet $\psi_{s,\tau}$ is defined to be the lesser of the raw wavelet length or the signal length, $N_\tau = \min (N_{t'},N_t)$.  The offset $\tau(t)$ is determined from either the center of the signal or the location of the cone-of-admissibility, and the algorithm to keep everything aligned gets a bit complicated: for $\tau' = \max (0 , \lfloor s \gamma - \lfloor N_t/2 )$ and $t' \in [- \lfloor s \gamma, \lfloor s \gamma]$ with duration $N_{t'}$, if $\tau \leq 0$ then $t' \rightarrow t'[1, \min (N_{t'},N_t)] + \tau'$, else $t' \rightarrow t'[\max(1,N_{t'}-N_t+1), N_{t'}] - \tau'$.  The end result is simply to truncate either edge of the wavelet as necessary, as shown in Fig.~\ref{fig:H}.  Then for the amplitude of the truncated wavelet $\psi_{s,\tau} = C_{s,\tau} \Phi^0_{s,\tau} \Theta^0_{s,\tau}$, with $C_{s,\tau} \equiv C_s / (\vert C_s \Phi^0_{s,\tau} \Theta^0_{s,\tau} \vert^2 s^2 / 2)^{1/2}$ we define the renormalized continuous wavelet transform (RCWT).  The procedure amounts to equalizing the norm of a truncated wavelet with that of a wavelet spanned entirely by the data record.

\begin{figure}[t]
\centerline{\psfig{file=\bw 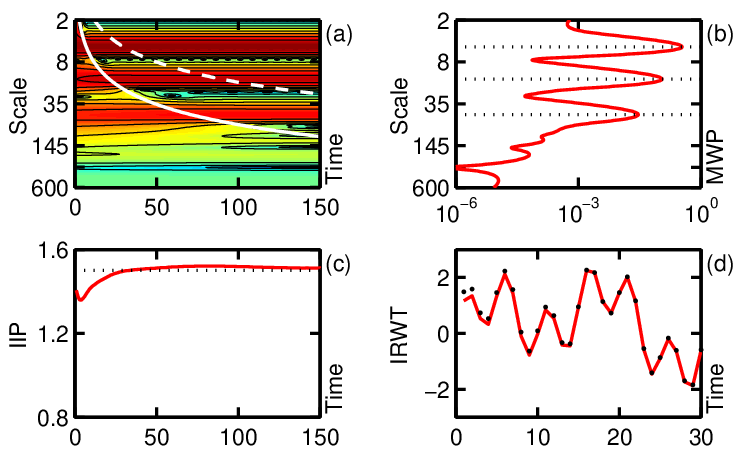}}
\vspace*{8pt}
\caption{\label{fig:C}
RCWT power spectral density (a), mean wavelet power (b), integrated instant power (c), and reconstruction (d) for test signal $y_1$.  Reconstruction is maintained until the edge-most sample locations. }
\end{figure}

Considering the same test signal as above, in Fig.~\ref{fig:C} we display the analysis using the RCWT; the reconstruction in (d) is noticeably improved, and the power estimation in (c) is not as affected near the signal edge.  The apparent increase in the IIP over the rms value represents we feel an aliasing in time, rather than scale, of the total power, as the mean discrepancy from the rms power is small.  In Table~\ref{tab:1} we display the ratio of the mean integrated power $\mrm{MIP} = N_t^{-1} \sum_t \sum_{s=2}^{s_{max}} \mrm{PSD} \Delta_s$ to the rms signal power $P_{rms} = 1.5$ for the CWT and RCWT, considering also an integration over scale which stops at the signal duration $s_{max} = N_t$ rather than $s_{max} = 2 N_t$.

\begin{table}[b]
\tbl{Ratio of mean integrated power to rms signal power for test signal $y_1$.
\label{tab:1} }
{\begin{tabular}{@{}ccccc@{}}
\toprule
PSD & \multicolumn{2}{c}{CWT} & \multicolumn{2}{c}{RCWT} \\ \colrule
$s_{max}$ & 600 & 300 & 600 & 300 \\
$\mrm{MIP} / P_{rms}$ &  0.95298 & 0.95242 & 0.99726 & 0.99561 \\
\botrule
\end{tabular} }
\end{table}

As pointed out by Frick {\it et al},\cite{Frick:1997426} wavelet truncation also affects the admissibility condition.  One commonly subtracts from $\Theta^0_{s,t} \equiv e^{i \omega_1 \eta}$ the DC component $\wt{\psi}_1(0) \propto e^{- \omega_1^2/2} \equiv d^c \sim 10^{-9}$ of the mother wavelet so that the zero mean wave becomes $\Theta_{s,t} = \Theta^0_{s,t} - d^c$.  For a truncated wavelet we take $d^c \rightarrow d^c_{s,t}$ to define the adaptive wavelet transform (AWT), where $d^c_{s,t} \equiv \langle \Theta^0_{s,\tau} \rangle = \sum_{t'} \Theta^0_{s,\tau} \Phi^0_{s,\tau} / \sum_{t'} \Phi^0_{s,\tau}$ is the weighted mean of the remaining wave $\Theta^0_{s,\tau}$.  Normalization as above with $C_{s,\tau}$ then defines the renormalized adaptive wavelet transform (RAWT) of Ref.~\refcite{rwj:astro01}.  In practice, we have found that the RCWT neglecting admissibility outperforms the RAWT by a small but noticeable margin: the troughs between peaks are slightly deeper, and the reconstruction is slightly better.  The reason, we feel, is that the adaptive admissibility condition alters the shape of the wavelet, hence its frequency response.

\section{Comparison to the Fourier Transform}
\label{dftcomp}
\begin{figure}[t]
\centerline{\psfig{file=\bw 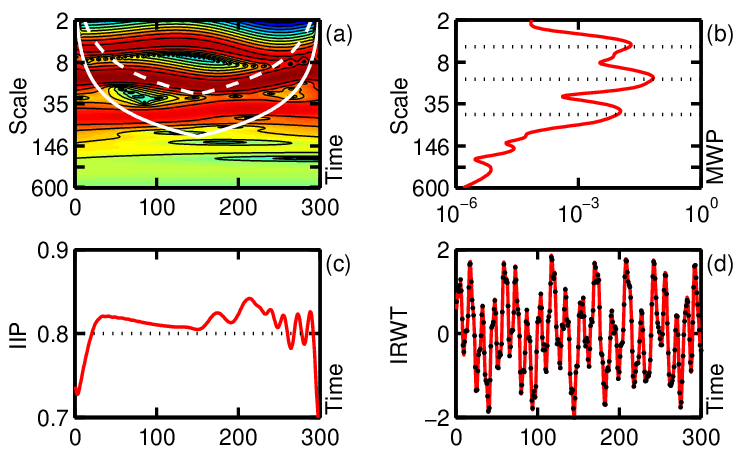}}
\vspace*{8pt}
\caption{\label{fig:D}
RCWT power spectral density (a), mean wavelet power (b), integrated instant renormalized power (c), and reconstruction (d) for test signal $y_2$ with components of time-varying periods around 5, 15, and 50 and squared amplitudes of .1, 1, and .5 respectively. }
\end{figure}

\begin{figure}[t]
\centerline{\psfig{file=\bw 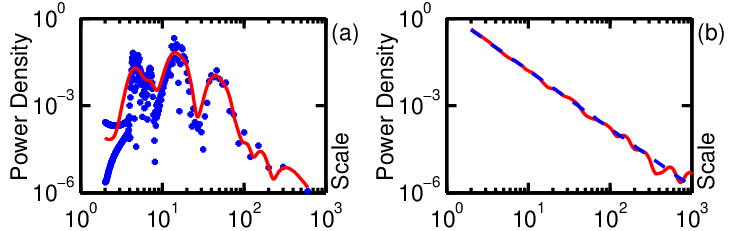}}
\vspace*{8pt}
\caption{\label{fig:E}
The one-sided Fourier periodogram $P_s$ ($*$) agrees with the renormalized MWP (solid) for signal $y_2$ as seen in (a).  The RCWT spectrum for white noise is shown in (b) compared to a line with logarithmic slope of -2 (dashed). }
\end{figure}

The hallmark of wavelet analysis is its ability to track signal components with periods that vary in time, in contrast to the Fourier transform whose basis functions span the entire data record.  Considering now a test signal $y_2$ of the same duration $N_t = 300$, for periods 5, 15, and 50 we adjust the squared amplitudes to be 0.1, 1, and 0.5 respectively (rms power of 0.8) and impose independent sinusoidal variation to the periods on the order of the duration.  In Fig.~\ref{fig:D} we display the RCWT analysis of such a signal.  The IIP in (c) again agrees with the rms power, and the reconstruction in (d) faithfully reproduces the signal.  Using the one-sided continuous Fourier transform (CFT)~\cite{Press-1992} \beq
\wt{y}_f = \sqrt{2} \sum_{t=1}^{N_t} y_t e^{i \pi f t / N_f} \Delta_t \;, \qquad y_t = \real \sqrt{2} \sum_{f=0}^{N_f} \wt{y}_f e^{-i \pi f t / N_f} \Delta_f \;,
\eeq evaluated at positive frequencies $f \Delta_f$ for $f \in [0, N_f]$ and $1/ \Delta_f = 2 N_f$ (recalling $\Delta_t \equiv 1$ and noting the two edge pixels have a width half that of the others), Plancherel's theorem for conservation of energy (or total power) is written $\sum_t \vert y_t \vert^2 \Delta_t = \sum_f \vert \wt{y}_f \vert^2 \Delta_f$, which when normalized by the duration $N_t$ gives the mean power of the signal.  To display the power distribution (periodogram) against an abscissa of scale $s = 1 / f$ as shown in Fig.~\ref{fig:E}(a), one must account for the integration measure\cite{Press-1992,Sivia-1996} so that $P_s = P_f \vert df/ds \vert = P_f / s^2$.  A power distribution that is constant in $f$, such as for white Gaussian noise, should appear against $s$ with a logarithmic slope of -2, and that is indeed what we find for the RCWT using a noise signal $y_3$ with duration 3000 as shown in (b).  Displayed in Table~\ref{tab:2} is the mean power for signals $y_2$ and $y_3$ evaluated for both the Fourier and Morlet transforms, using a trapezoidal quadrature for the integration of the wavelet power.

\begin{table}[t]
\tbl{Comparison of CFT and RCWT mean signal power.
\label{tab:2} }
{\begin{tabular}{@{}ccccc@{}}
\toprule
signal & $N_t^{-1} \sum_t \vert y_t \vert^2$ & $N_t^{-1} \sum_f \vert \wt{y}_f \vert^2 \Delta_f$ & $N_t^{-1} \sum_s P_s \Delta_s$ & $\sum_s \mrm{MWP}_s \Delta_s$ \\ \colrule
$y_2$ & 0.79793 &  0.79793  & 0.79793 &  0.79758 \\
$y_3$ & 0.81972 &  0.81972  & 0.81972 &  0.82067 \\
\botrule
\end{tabular} }
\end{table}

\newpage

\section{Enhancement}
\label{enhance}
With the transform now responding like a theoretical apparatus for measuring a signal's power spectral density, one may apply the techniques of resolution enhancement common in the analysis of experimental data.\cite{Sivia-1996,Bretthorst-1988}  Our approach considers the RCWT algorithm as providing a mathematical model for some spectroscopic device of finite resolution, so that the power of an input signal is distributed according to the device's resolving capabilities as measured by its point spread function or response matrix.  A device with infinitesimal resolution has a point spread function equal to the identity so that a Dirac distribution upon input is mirrored on output.  A finite resolution gives the output distribution a width which results from the convolution of the response with the input spectrum.  An experimental device is calibrated by determining the point spread function for a collection of known input distributions so that the measurements of an unknown signal may be deconvolved to yield the best estimate of the spectrum.  We can follow the same procedure using our theoretical apparatus.

\begin{figure}[b]
\centerline{\psfig{file=\bw 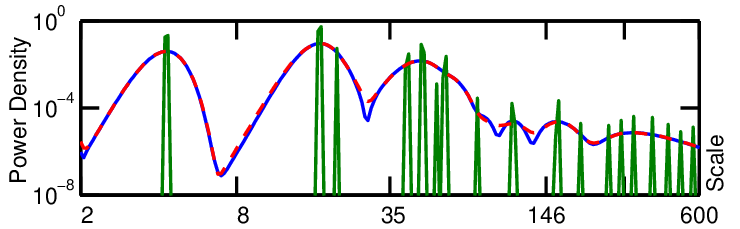}}
\vspace*{8pt}
\caption{\label{fig:F}
EIP (spikes) and IWP (solid) at the midpoint of the signal duration using a tolerance of $10^{-8}$. The reconstructed enhanced power (dashed) slightly exceeds the original IWP at the trough locations. }
\end{figure}

First one writes the point spread function as the response matrix $R(s,s')$ defined by the integrated power of a wavelet of scale $s$ convoluted with a signal component of period $s'$.  The point spread function represents the spectral leakage of the transform.  For this analysis we take the signal components to be cosine functions for the duration of the wavelet, \beq
R(s,s') = 2 \biggl \vert \sum_{t'} \bar{\psi}_{s,0}(t') \cos(2 \pi t' / s') \biggr \vert^2 \;,
\eeq where the factor of 2 accounts for a signal of unit power.  We note that here one is making an assumption on the form of the underlying signal elements whose composition represents the original signal, and that currently our implementation of the enhancement does not account for wavelet truncation, which would require an offset dependent response matrix $R_t(s,s')$.  Then, for each IWP in the PSD, the enhanced instant power EIP is the solution to the equation \beq
0 = \sum_s \biggl [ \sum_{s'} R(s,s') {\rm EIP}_t(s') \Delta_{s'} - {\rm IWP}_t(s) \biggr ]^2 \Delta_s \;,
\eeq found in a least-squares sense with non-negativity constraints.  Note that it is the redundancy in scale of the CWT which provides the resolution enhancement of the EIP.  The effect is to replace broad peaks in the IWP with sharp spikes at the scale of the corresponding signal component, as shown in Fig.~\ref{fig:F} for the IWP at the midpoint of the duration of the signal $y_2$.  The reconstructed enhanced power $\mrm{REP} = R \times \mrm{EIP}$ (dashed) differs slightly from the original IWP (solid) as no constraint has been placed on preserving the norm.  In general, one's wavelet response may extend beyond one's region of calculation for signal periods near either cutoff, and the enhancement procedure is capable of recapturing the lost (uncomputed) power.  If one's application indicates the signal is bandwidth limited to that region well within the cone-of-influence yet far from the Nyquist scale, then enforcement of a norm-preserving constraint during the minimization is suggested.  The enhanced power spectral density $\mrm{EPSD}(s,t)$ is then defined simply as the collection of enhanced instant powers.

\begin{figure}[t]
\centerline{\psfig{file=\bw 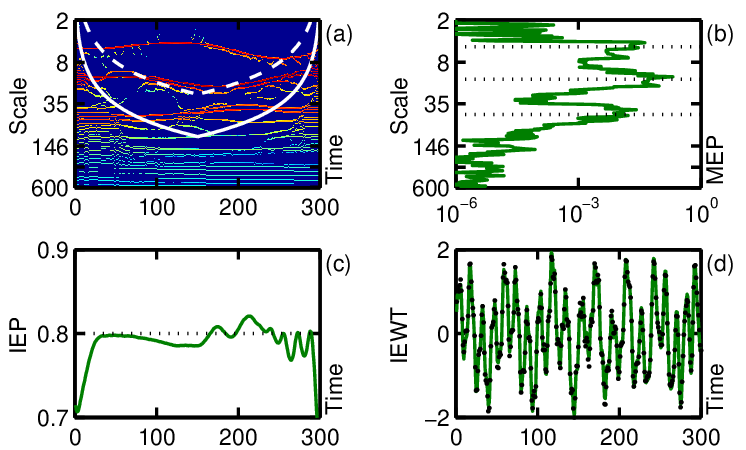}}
\vspace*{8pt}
\caption{\label{fig:G}
Enhanced power spectral density (a), mean enhanced power (b), integrated enhanced power (c), and reconstruction (d) for test signal $y_2$ with components of time-varying periods around 5, 15, and 50. }
\end{figure}

In Fig.~\ref{fig:G} we show the EPSD for test signal $y_2$, as well as the mean enhanced power MEP and integrated enhanced power IEP, which is less than the previous IIP by a small amount.  The variation in scale of the signal periods is well-resolved within the cone-of-influence, and a video scanning through the EIPs is available as an online supplement.  Reconstruction from the EPSD is not yet well-defined; however, one may attempt a reconstruction using the phase of the RCWT and the original renormalized basis as shown in (d), which might not be perfect but does faithfully represent the original signal after normalization by an arbitrary constant.  For signal components within the cone-of-influence, the EPSD provides the maximum resolution in scale available from the RCWT.

\section{Conclusions}
\label{concl}
The utility of these algorithms\footnote{Available as the AlphaWavelet Toolbox at http://www.alphawaveresearch.com.} should be apparent to anyone familiar with one dimensional data analysis and power spectrum estimation.  Extension of the renormalization prescription to multi-dimensional wavelet analysis is straightforward; less so for the enhancement procedure.  The symmetric normalization adopted here returns a power spectral density which behaves exactly as it should, with margins that give the mean and instant power of a signal in physical units, and provides nearly perfect reconstruction without the introduction of an arbitrary factor.  The mean wavelet power agrees with the discrete Fourier transform on the distribution of power for a signal, and the integrated instant power agrees with the rms power of the signal components.

We have recently\cite{rwj:jgr02,rwj:astro01} applied algorithms very similar to the ones presented here to solar analysis.  In that work, by addressing the power spectral density of the historical sunspot record, a relation is found between the level of solar magnetic activity and the temperature observed in central England.  What makes that comparison possible is the replacement of the yearly index with one for solar activity as evaluated by the integrated instant power.  One also may consider its application to signal encoding, manipulation, and compression, providing an alternate basis for reconstruction.  For temporally resolved power spectrum estimation, the symmetric wavelet transform has become quite a useful tool indeed.

In summary, the continuous wavelet transform using the Morlet basis may be normalized to account for the response at negative scales, resulting in a symmetric forward and inverse transform pair with nearly perfect reconstruction.  It may then be renormalized to account for wavelet truncation by keeping a constant wavelet power for each scale, where neglecting the admissibility condition results in better performance for data analysis, extending the useful range beyond the cone-of-influence.  By equalizing the wavelet norm, the renormalized transform allows for a quantitative estimate of the power spectral density in physical units for the duration of the data record.  Deconvolution with the wavelet response matrix then yields the enhanced power spectral density, providing the maximum resolution in scale of the harmonic content carried by a signal.



\begin{thebibliography}{10}

\addtolength{\baselineskip}{-.016\baselineskip} 

\bibitem{Morlet:1984}
P.~Goupillaud, A.~Grossman, and J.~Morlet, Cycle-octave and related transforms
  in seismic signal analysis, \emph{Geoexploration} \textbf{23} (1984) 85--102.

\bibitem{kaiser:1990}
G.~Kaiser, \emph{Quantum Physics, Relativity and Complex Spacetime: Towards a
  New Synthesis} (North-Holland, 1990).

\bibitem{kaiser:1994}
G.~Kaiser, \emph{A Friendly Guide to Wavelets} (Birkhauser, 1994).

\bibitem{perel-3441P}
M.~V. Perel and M.~S. Sidorenko, New physical wavelet `{G}aussian wave packet',
  \emph{J. Phys. A-Math. Theor.} \textbf{40} (2007)
  3441--3461.

\bibitem{perel-5211P}
M.~V. Perel and M.~S. Sidorenko, Wavelet-based integral representation for
  solutions of the wave equation, \emph{J. Phys. A-Math. Theor.} \textbf{42} (2009) 375211.

\bibitem{lewalle-131}
J.~Lewalle, Field reconstruction from single scale continuous wavelet
  coefficients, \emph{Int. J. Wavelets Multi.} \textbf{7} (2009) 131--142.

\bibitem{karak-313}
S.~Karakaz'yan, M.~Skopina, and M.~Tchobanou, Symmetric multivariate wavelets,
  \emph{Int. J. Wavelets Multi.} \textbf{7} (2009) 313--340.

\bibitem{echera-41}
M.~P.~S. Echer, E.~Echer, D.~J.~R. Nordemann, and N.~R. Rigozo,
  Multi-resolution analysis of global surface air temperature and solar
  activity relationship, \emph{J. Atmos. Sol.-Terr. Phy.} \textbf{71} (2009) 41--44.

\bibitem{fligge-313}
M.~{Fligge}, S.~K. {Solanki}, and J.~{Beer}, {Determination of solar cycle
  length variations using the continuous wavelet transform}, \emph{Astron.
  Astrophys.} \textbf{346} (1999) 313--321.

\bibitem{Foster:1996}
G.~Foster, Wavelets for period analysis of unevenly samples time series,
  \emph{Astron. J.} \textbf{112} (1996) 1709--1729.

\bibitem{Frick:1997426}
P.~Frick, S.~L. Baliunas, D.~Galyagin, D.~Sokoloff, and W.~Soon, Wavelet
  analysis of stellar chromospheric activity variations, \emph{Astrophys. J.}
  \textbf{483} (1997) 426--434.

\bibitem{Frick:1997670F}
P.~{Frick}, D.~{Galyagin}, D.~V. {Hoyt}, E.~{Nesme-Ribes}, K.~H. {Schatten},
  D.~{Sokoloff}, and V.~{Zakharov}, {Wavelet analysis of solar activity
  recorded by sunspot groups}, \emph{Astron. Astrophys.} \textbf{328} (1997)
  670--681.

\bibitem{ligao-181}
K.~J. {Li}, P.~X. {Gao}, and T.~W. {Su}, The {S}chwabe and {G}leissberg periods
  in the {W}olf sunspot numbers and the group sunspot numbers, \emph{Sol. Phys.}
  \textbf{229} (2005) 181--198.

\bibitem{piscaron-1661}
P.~Piscaronoft, J.~Kalvov\'{a}, and R.~Br\'{a}zdil, Cycles and trends in the
  {C}zech temperature series using wavelet transforms, \emph{Int. J. Climatol.}
  \textbf{24} (2004) 1661--1670.

\bibitem{polygian-725}
J.~{Polygiannakis}, P.~{Preka-Papadema}, and X.~{Moussas}, {On signal-noise
  decomposition of time-series using the continuous wavelet transform:
  {A}pplication to sunspot index}, \emph{Mon. Not. R. Astr. Soc.} \textbf{343}
  (2003) 725--734.

\bibitem{sweldens98lifting}
W.~Sweldens, The lifting scheme: {A} construction of second generation
  wavelets, \emph{SIAM J. Math. Anal.} \textbf{29} (1998)
  511--546.

\bibitem{torrence:98}
C.~Torrence and G.~P. Compo, A practical guide to wavelet analysis,
  \emph{B. Am. Meteorol. Soc.} \textbf{79} (1998)
  61--78.

\bibitem{weber-917}
S.~L. {Weber}, {A timescale analysis of the Northern Hemisphere temperature
  response to volcanic and solar forcing}, \emph{Clim. Past}
  \textbf{1} (2005) 9--17.

\bibitem{liu-2093}
Y.~Liu, X.~S. Liang, and R.~H. Weisberg, Rectification of the bias in the
  wavelet power spectrum, \emph{J. Atmos. Oceanic Technol.} \textbf{24} (2007)
  2093--2102.

\bibitem{rwj:jgr02}
R.~W. {Johnson}, Enhanced wavelet analysis of solar magnetic activity with
  comparison to global temperature and the Central England Temperature record,
  \emph{J. Geophys. Res.-Space} \textbf{114} (2009)
  A05105.

\bibitem{rwj:astro01}
R.~W. {Johnson}, Edge adapted wavelets, solar magnetic activity, and climate
  change, \emph{Astrophys. Space Sci.} \textbf{326} (2010) 181--189.

\bibitem{Press-1992}
W.~H. {Press}, S.~A. {Teukolsky}, W.~T. {Vetterling}, and B.~P. {Flannery},
  \emph{Numerical Recipes} (Cambridge University Press, 1992).

\bibitem{Sivia-1996}
D.~S. Sivia, \emph{Data Analysis: A Bayesian Tutorial} (Oxford University
  Press, 1996).

\bibitem{Bretthorst-1988}
G.~L. Bretthorst, \emph{Bayesian Spectrum Analysis and Parameter Estimation}
  (Springer-Verlag, 1988).

\end{thebibliography}

\end{document}